\documentclass[%
 reprint,
superscriptaddress,
showpacs,showkeys,preprintnumbers,
 amsmath,amssymb,
 aps,
]{revtex4-1}

\usepackage{graphicx}
\usepackage{dcolumn}
\usepackage{bm}
\usepackage{here}
\usepackage[dvips]{color}
\usepackage{multirow}
\usepackage{ulem}

\makeatletter
  \def\erf{\mathop{\operator@font erf}\nolimits}
  \def\erfc{\mathop{\operator@font erfc}\nolimits}
  \def\Erf{\mathop{\operator@font Erf}\nolimits}
  \def\Shi{\mathop{\operator@font Shi}\nolimits}
  \def\Chi{\mathop{\operator@font Chi}\nolimits}
  \def\Ei{\mathop{\operator@font Ei}\nolimits}
  \def\cosec{\mathop{\operator@font cosec}\nolimits}
  \def\sech{\mathop{\operator@font sech}\nolimits}
  \def\cosech{\mathop{\operator@font cosech}\nolimits}
  \newcommand\hypgeo[2]{{}_{#1}{\operator@font F}_{#2}}
  \def\Re{\mathop{\operator@font Re}\nolimits}
  \def\Im{\mathop{\operator@font Im}\nolimits}
\makeatother

\begin{document}


\title{Angular distribution of $\gamma$-rays from a neutron-induced $p$-wave resonance of $^{132}$Xe}

\def\affNagoya{Nagoya University, Furocho, Chikusa, Nagoya 464-8602, Japan}
\def\affKyushu{Kyushu University, 744 Motooka, Nishi, Fukuoka 819-0395, Japan}
\def\affJAEA{Japan Atomic Energy Agency, 2-1 Shirane, Tokai 319-1195, Japan}
\def\affTokyoTech{Tokyo Institute of Technology, Meguro, Tokyo 152-8551, Japan}
\def\affRCNP{Osaka University, Ibaraki, Osaka 567-0047, Japan}
\def\affIndiana{Indiana University, Bloomington, IN 47401, USA}
\def\affLosAlamos{Los Alamos National Laboratory, Los Alamos, NM 87545, USA}
\def\affKEK{High Energy Accelerator Research Organization,1-1 Oho, Tsukuba, Ibaraki 305-0801, Japan}
\def\affTohoku{Tohoku University, 41 Kawauchi, Aoba-ku, Sendai, 980-8576 Japan}

\author{T.~Okudaira}
\affiliation{\affNagoya}

\author{Y.~Tani}
\affiliation{\affTokyoTech}

\author{S.~Endo}
\affiliation{\affNagoya}
\affiliation{\affJAEA}

\author{J.~Doskow}
\affiliation{\affIndiana}

\author{H.~Fujioka}
\affiliation{\affTokyoTech}

\author{K.~Hirota}
\thanks{Present Address: \affKEK}
\affiliation{\affNagoya}

\author{K.~Kameda}
\affiliation{\affTokyoTech}

\author{A.~Kimura}
\affiliation{\affJAEA}

\author{M.~Kitaguchi}
\affiliation{\affNagoya}

\author{M.~Luxnat}
\affiliation{\affIndiana}

\author{K.~Sakai}
\affiliation{\affJAEA}

\author{D.~Schaper}
\affiliation{\affLosAlamos}

\author{T.~Shima}
\affiliation{\affRCNP}

\author{H.~M.~Shimizu}
\affiliation{\affNagoya}

\author{W.~M.~Snow}
\affiliation{\affIndiana}

\author{S.~Takada}
\thanks{Present Address: \affTohoku}
\affiliation{\affKyushu}

\author{T.~Yamamoto}
\affiliation{\affNagoya}

\author{H.~Yoshikawa}
\affiliation{\affRCNP}

\author{T.~Yoshioka}
\affiliation{\affKyushu}


\date{\today}


\begin{abstract}


A neutron-energy dependent angular distribution was measured for individual $\gamma$-rays from the 3.2~eV $p$-wave resonance of $^{131}$Xe+$n$,  that shows enhanced parity violation owing to a mixing between $s$- and $p$-wave amplitudes. The $\gamma$-ray transitions from the $p$-wave resonance were identified, and the angular distribution with respect to the neutron momentum was evaluated as a function of the neutron energy for 7132~keV $\gamma$-rays, which correspond to a transition to the 1807~keV excited state of $^{132}$Xe. The angular distribution is considered to originate from the interference between $s$- and $p$-wave amplitudes, and will provide a basis for a quantitative understanding of the enhancement mechanism of the fundamental parity violation in compound nuclei.

\end{abstract}

\pacs{13.75.Cs, 
21.10.Hw,
21.10.Jx,
21.10.Re,
23.20.En,
24.30.Gd,
24.80.+y,
25.40.Fq,
25.70.Gh,
27.60.+j,
29.30.Kv
}
\keywords{compound nuclei,
partial wave interference,
neutron radiative capture reaction}
\maketitle


\section{Introduction}
Parity odd asymmetries on the order of $10^{-6}$ have been observed in the helicity dependence of the scattering cross section in nucleon-nucleon scattering experiments~\cite{pot74,yua86,ade85}. This tiny parity violation is understood to come from the weak amplitude in the hadronic interaction, which is on the order of $10^{-6}-10^{-7}$ times the strong amplitude. In the neutron absorption reaction of nuclei with a mass number of around 100, large parity violations with sizes of up to 10\% have been observed on $p$-wave resonances located on the tail of $s$-wave resonances. The fundamental parity violation in the nucleon-nucleon interaction is enhanced by the mixing between $s$- and $p$-wave amplitudes of compound nuclei, referred to as the $s$-$p$ mixing model~\cite{sus82,sus82E}.\\
The interference terms between $s$- and $p$-wave amplitudes introduced by the $s$-$p$ mixing model implies several correlations between neutron momentum, neutron polarization, $\gamma$-ray momentum, and $\gamma$-ray polarization at the $p$-wave resonance~\cite{fla85}. These correlations provide important information needed to understand the enhancement mechanism of the parity violation in compound nuclei (e.g. partial neutron width and weak matrix element). They are measured as a neutron-energy dependent angular distributions of emitted $\gamma$-rays from $p$-wave resonances. Non-uniform angular distributions were observed in reactions of $^{139}$La($n$,$\gamma$)$^{140}$La*~\cite{okuda21},  $^{139}$La($n$,$\gamma$)$^{140}$La$_{\rm{g.s.}}$~\cite{okuda18}, $^{139}$La($\vec{{n}}$,$\gamma$)$^{140}$La$_{\rm{g.s.}}$~\cite{yama20} and $^{117}$Sn($n$,$\gamma$)$^{118}$Sn$_{\rm{g.s.}}$~\cite{koga22},$^{117}$Sn($\vec{{n}}$,$\gamma$)$^{118}$Sn$_{\rm{g.s.}}$~\cite{endo22}. Fundamental time reversal violation can be enhanced by the same mechanism, and these nuclei can be utilized to search for T-violation in the nucleon-nucleon interaction by measuring a forward scattering amplitude with polarized neutrons and a polarized or aligned nuclear target~\cite{gud92}.\\
The 3.2~eV $p$-wave resonance of $^{131}$Xe+$n$ is an important compound state in the study of the $s$-$p$ mixing model and the search for new physics. A very large parity-odd longitudinal asymmetry with a size of (4.3 $\pm$ 0.2)\% was observed in the 3.2~eV $p$-wave resonance of $^{131}$Xe+$n$ with a neutron transmission measurement~\cite{szy96}. This parity violation is the second largest for $p$-wave resonances in the eV region, where high neutron intensity can be obtained at spallation sources, after (9.56 $\pm$ 0.35)\% for $^{139}$La~\cite{LANL91}. A $^{131}$Xe nuclear polarization of $7.6 \pm 1.5$\% using spin exchange optical method (SEOP) was also recently achieved~\cite{mol21}.\\
The ($n$, $\gamma$) measurement in Ref.~\cite{skoy96} suggested that the 3.2~eV $p$-wave resonance originates from $^{131}$Xe+$n$, but no detailed study of the $p$-wave resonance with $\gamma$-ray measurements has been performed since then. The 3.2~eV $p$-wave resonance has a very small cross section, which requires a large neutron beam intensity and large solid angle of $\gamma$-ray detectors. Also, there is a very large 14.4~eV $s$-wave resonance in $^{131}$Xe+$n$ ($2.0\times10^4$ barn) near the $p$-wave resonance. $\gamma$-rays from a very large 14.4~eV $s$-wave resonance saturate $\gamma$-ray detectors, and thus it is difficult to increase the neutron beam intensity.
In this paper, we report the angular distribution of the $\gamma$-rays from the 3.2~eV $p$-wave resonance with a high statistics by employing an enriched $^{131}$Xe filter upstream of the detector array, that selectively absorbs incident neutrons around 14.4~eV to avoid the detector saturation around the $p$-wave resonance.

\section{Experiment}
\subsection{Experimental Setup}
\begin{figure}[htbp]
	\centering
	\includegraphics[width=0.95\linewidth]{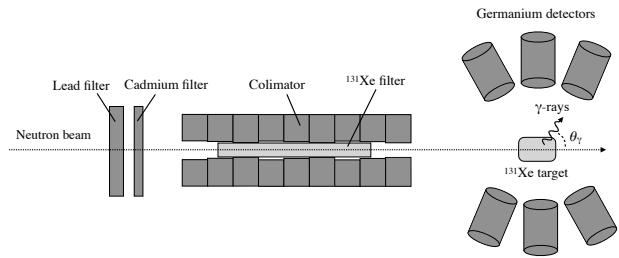}
	\caption[]{
	Experimental setup.  
	}
	\label{setup}
\end{figure}

The $^{131}$Xe($n$, $\gamma$)$^{132}$Xe reaction was measured with a pulsed epithermal neutron beam at Accurate Neutron-Nucleus Reaction Measurement Instrument (ANNRI) beamline of the Material and Life Science Experimental Facility (MLF) of the Japan Proton Accelerator Research Complex (J-PARC). The neutron induced prompt $\gamma$-rays were measured with 22 germanium detectors. The germanium detectors are oriented at angles from 36$^\circ$ to 144$^\circ$ with respect to the incident neutron beam direction~\cite{kim12}. The neutron beam was collimated to 22~mm using epithermal neutron collimators installed upstream of the germanium detectors. The detailed description of the beamline is given in Ref.~\cite{okuda18}.\\
The experimental setup is shown in Fig.~\ref{setup}. The nuclear target was 84.4\% enriched $^{131}$Xe gas. The Xe gas was encapsulated into a cylindrical cell with a pressure of 0.43 MPa, a diameter of 49 mm and a thickness of 76 mm. The target was placed at the detector center, and the distance from the moderator surface to the Xe target was 21.5~m.  The $^{131}$Xe filter was an 84.4\% enriched $^{131}$Xe gas encapsulated with a pressure of 0.79~MPa at room temperature in a cylindrical aluminum cell with an inner diameter of 23~mm and a thickness of 500~mm. It was installed inside the collimator, and the distance from the moderator surface to the center of the $^{131}$Xe filter was 20.0~m. Most of incident neutrons around the $s$-wave resonance were absorbed by the $^{131}$Xe filter, whereas the neutrons around 3.2 eV passed through it because of the small cross section of the $p$-wave resonance. A cadmium filter with 1~mm thickness was installed to absorb thermal neutrons. As we shall see later, $\gamma$-rays of $^{56}$Fe($n$,$\gamma$)$^{57}$Fe reactions from the upstream of the beamline was the main background in this experiment. A lead filter was used to suppress $\gamma$-ray background, moderately sacrificing  the neutron beam intensity. Measurements were performed under two different $\gamma$-ray backgrounds and neutron beam intensities with 37.5~mm-thick and 50~mm-thick  lead filters. The proton beam power was 600~kW and the total measurement time was 227 hours. 

\subsection{Measurement}
\begin{figure}[htbp]
	\centering
	\includegraphics[width=0.95\linewidth]{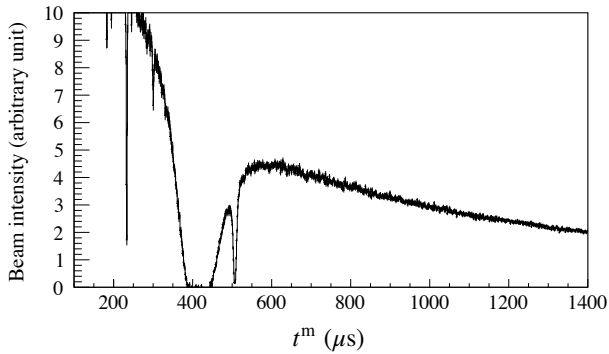}
	\caption[]{
	Relative incident neutron spectrum as a function of $t^{\rm m}$ measured with the boron target. The neutron intensity around 410~$\rm{\mu s}$ and 500~$\rm{\mu s}$, which correspond to the peak energies of the 14.4~eV $s$-wave resonance of $^{131}$Xe+$n$ and 9.66~eV $s$-wave resonance of $^{129}$Xe+$n$, respectively, was reduced owing to the neutron absorber.  
	}
	\label{Beam}
\end{figure}

\begin{figure}[htbp]
	\centering
	\includegraphics[width=0.95\linewidth]{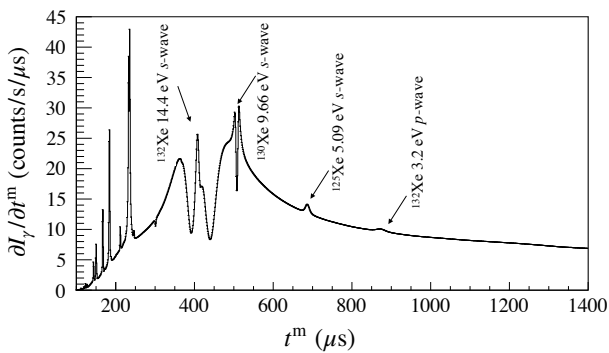}
	\caption[]{
	$\gamma$-ray counts as a function of $t^{\rm m}$ with the $^{131}$Xe target. The peak at 410~$\rm{\mu s}$ is considered to originate from neutrons passing through the gap between the collimator and the $^{131}$Xe filter.
	}
	\label{TOF}
\end{figure}

We measured the energy deposition and the arrival time of the capture $\gamma$-rays event-by-event in a list mode.
The energy deposition of $\gamma$-rays in the germanium detector $E_{\gamma}^{\rm m}$ is determined from the pulse height. \\ \\ The time-of-flight of the incident neutrons was obtained using the arrival time of the $\gamma$-rays $t^{\rm m}$ from intense $\gamma$-flash emitted during the proton spallation reactions in the target.  The corresponding neutron energy $E_{n}^{\rm m}$ is calculated using $t^{\rm m}$ and the known beamline geometry. The neutron energy in the center-of-mass system $E_{n}$ is defined as well. The total number of $\gamma$-ray events detected in the experiment are denoted as $I_{\gamma}$. A 2-dimensional histogram corresponding to $\partial^2I_{\gamma}/\partial t^{\rm m}\partial E^{\rm m}_\gamma$ was obtained for each germanium crystal. The detailed definitions of variables are described in Ref.~\cite{okuda18}. \\
An incident beam spectrum for $t^{\rm m}$, obtained from a measurement of the 477.6~keV $\gamma$-rays in the neutron absorption reaction of ${}^{10}$B with an enriched ${}^{10}$B target, is shown in Fig.~\ref{Beam}. Figure~\ref{Beam} shows that the incident neutrons were selectively suppressed around the $s$-wave resonances of Xe+$n$ due to neutron absorption reactions in the $^{131}$Xe filter. \\
The histogram of $\partial I_{\gamma}/\partial t^{\rm m}$ measured with the$^{131}$Xe target is shown in Fig.~\ref{TOF}. The small peak at $t^{\rm m}\sim880~\mu\rm{s}$ is the 3.2~eV $p$-wave resonance of $^{131}$Xe+$n$, and the peak at $t^{\rm m}\sim410~\mu\rm{s}$ corresponds to the 14.4~eV $s$-wave resonance of $^{131}$Xe+$n$, which is strongly suppressed by the $^{131}$Xe filter. We corrected for 10\% and 5\% of the total $\gamma$-ray signals overlapped in time in the measurements with a 37.5~mm and 50~mm lead filter, respectively~\cite{okuda18}.

\begin{figure}[htbp]
	\centering
	\includegraphics[width=0.95\linewidth]{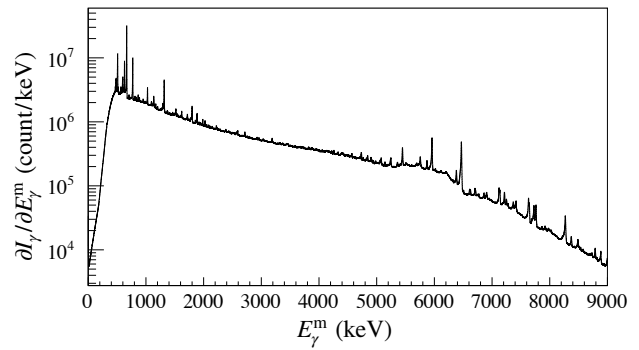}
	\caption[]{
	$\gamma$-ray spectrum defined as $\partial I_{\gamma}/\partial E_{\gamma}^{\rm m}$.
	}
	\label{Gamma}
\end{figure}

\begin{figure}[htb]
	\centering
	\includegraphics[width=0.95\linewidth]{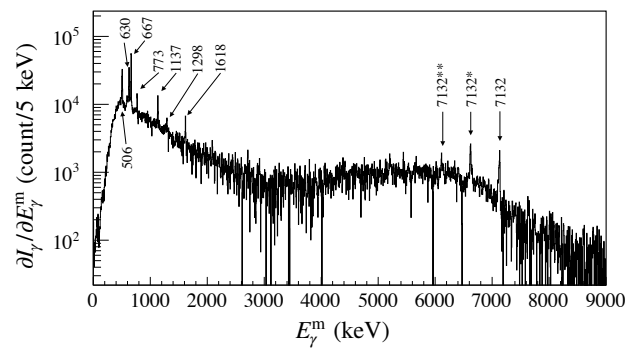}
	\caption[]{
	$\gamma$-ray spectrum gated with the 3.2~eV $p$-wave component. The $\gamma$-ray energies of each peak are also shown. Single and double asterisks indicate single- and double- escape peaks, respectively. The origin of the $\gamma$-ray peak at 1618~keV was not identified in this study. The detection threshold was around 500~keV}.
	\label{Gamma_p}
\end{figure}

\begin{figure}[htbp]
	\centering
	\includegraphics[width=0.9\linewidth]{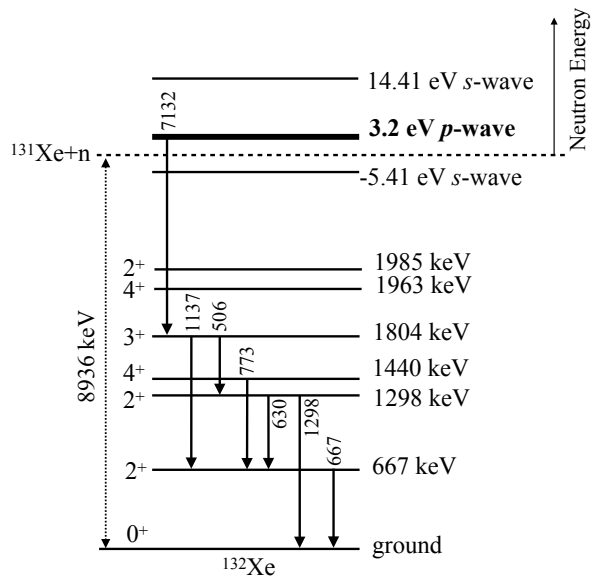}
	\caption[]{
	Decay scheme from the 3.2~eV $p$-wave resonance of $^{131}$Xe+$n$ to $^{132}$Xe. The dashed line shows separation energy of $^{131}$Xe+$n$. The excited states less than 2000~keV are depicted.
	}
	\label{Transition}
\end{figure}

\begin{figure}[htbp]
	\centering
	\includegraphics[width=0.9\linewidth]{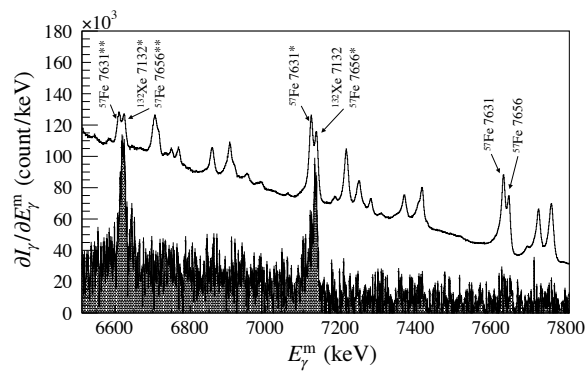}
	\caption[]{
	Expanded $\gamma$-ray spectra around the 7132~keV $\gamma$-ray peak. The black line and shaded area denote the histograms integrated for $t^{\rm m}>0~\mu\rm{s}$ and gated with the $p$-wave resonance, respectively. Single and double asterisks indicate single- and double- escape peaks, respectively. The single and double escape peaks from $^{56}$Fe($n$,${\gamma}$)$^{57}$Fe overlapp with the full absorption and single escape peaks of the 7132~keV $\gamma$-rays from $^{131}$Xe($n$,${\gamma}$)$^{132}$Xe. The histogram gated with the $p$-wave resonance is scaled by a factor of 200.} 
	\label{ExpandedGamma}
\end{figure}

\begin{table*}[htbp]
\begin{center}
	\begin{tabular}{c||c|c|c|c|c|c||c|c}
	\multirow{2}{*}{$r$} & 
	\multicolumn{6}{c||}{published values} &
	\multicolumn{2}{c}{this work} \\
	\cline{2-9}
	&
	$E_r\,[{\rm eV}]$ & 
	$J_r$ & 
	$l_r$ & 
	$\Gamma^{\gamma}_{ r}\,[{\rm meV}]$ &
	$2g_{ r}\Gamma^{n}_{ r}\,[{\rm meV}]$&
	 $2g_{ r}\Gamma_r^{{\rm n}l_r}\,[{\rm meV}]$ &
	$E_r\,[{\rm eV}]$ & 
	$\Gamma^{\gamma}_{ r}\,[{\rm meV}]$ \\
	\hline
	$1$ & $-5.41$ & $(1)$ & $$ &$123.7$ & & $2.93$ & \\
	$2$ & $3.2\pm0.3$ & & $1$ & $$ &  $(3.2 \pm0.3 )\times 10^{-5}$& &
		$3.20\pm 0.01$ & $174\pm 7$ \\
	$3$ & $14.41\pm0.014$ &(2) & $0$ & $93.5\pm6.2$ &$268\pm7$& &
		$$ & $$ \\
	\end{tabular}
	\caption{
	The resonance parameters of $^{131}$Xe+$n$ for low energy neutrons. The resonance parameters $E_r$, $J_r$, $l_r$, $\Gamma_r^\gamma$, $g_r$ ,and $\Gamma_r^n$ are resonance energy, total angular momentum, orbital angular momentum, $\gamma$ width, {\it{g}}-factor and neutron width, respectively. Parameter $r$ denotes the resonance number. The spin and parity of $^{131}$Xe are $3/2^+$, and therefore the $p$-wave resonance has negative parity. Published values are taken from Ref.~\cite{mughabghab}. 
	}
	
	\label{resopara}
\end{center}	
\end{table*}
Figure~\ref{Gamma} shows a $\gamma$-ray spectrum defined as $\partial I_{\gamma}/\partial E_{\gamma}^{\rm m}$, which is integrated for $t^{\rm m}<0~\mu\rm{s}$. The $\gamma$-ray spectrum gated with 2.98~eV $<E^{\rm{m}}_{\rm{n}}<$ 3.42~eV, which corresponds to the energy range of the $p$-wave resonance, is also shown in Fig.~\ref{Gamma_p}. Note that the $\gamma$-rays from the $s$-wave component and other backgrounds were subtracted in Fig.~\ref{Gamma_p} by estimating the $s$-wave and backgrounds components with a third polynomial fit in the histogram of $\partial I_{\gamma}/\partial t^{\rm m}$. The transitions from the $p$-wave resonance were identified from Fig.~\ref{Gamma_p} for the first time and the decay scheme for the 3.2~eV $p$-wave resonance was obtained as shown in Fig.~\ref{Transition}. The energy levels of $^{132}$Xe were taken from Ref.~\cite{NNDC}. The 7132~keV $\gamma$-ray, which corresponds to the transition to 1804~keV excited state of $^{132}$Xe, was observed as a direct transition from the 3.2~eV $p$-wave resonance. The 7132 keV $\gamma$-ray had not been found in previous studies, which focused on the $s$-wave component and the thermal region.\cite{PhysRevC.3.1678, NSR1971GR28, Hamada_1988}.\\
The full absorption peak and the single escape peak of the 7132~keV $\gamma$-rays were used in the study of the angular distribution of $\gamma$-rays. The expanded $\gamma$-ray spectra integrated for $t^{\rm m}>0~\mu\rm{s}$ and gated with the $p$-wave resonance are shown in Fig.~\ref{ExpandedGamma}. The main background for this study comes from the single and the double escape peaks of 7631~keV and 7646~keV $\gamma$-rays of the $^{56}$Fe($n$,$\gamma$)$^{57}$Fe reaction, which arises from iron material used in the beamline instrument. These $\gamma$-ray peaks from $^{56}$Fe+$n$ cannot be isolated in the $\gamma$-ray spectrum because the peaks completely overlap with the full absorption and single escape peaks of 7132~keV $\gamma$-rays as shown in Fig.~\ref{ExpandedGamma}. The background subtraction was performed in the histogram of $\partial I_{\gamma}/\partial t^{\rm m}$ for the evaluation of the angular distribution, which is described in subsection C.\\


\begin{figure}[htbp]
	\centering
	\includegraphics[width=0.9\linewidth]{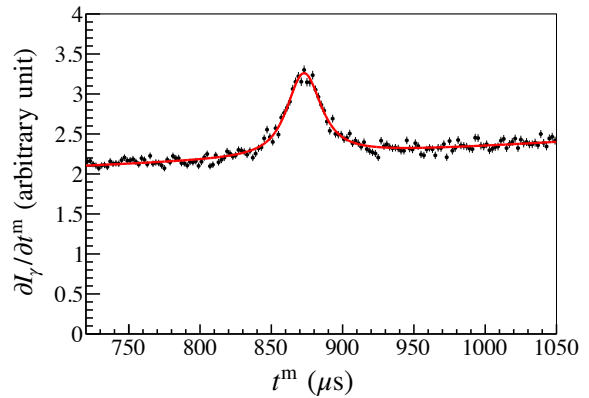}
	\caption[]{
	Histogram of $\partial I_{\gamma}/\partial t^{\rm m}$ gated with the full absorption and single escape peaks of 7132~keV $\gamma$-rays. The best fit is shown in the solid line. 
	}
	\label{Fitreso}
\end{figure}

The resonance parameters of the $p$-wave resonance were determined with the histogram of $\partial I_{\gamma}/\partial t^{\rm m}$ gated with the full absorption and single escape peaks of 7132~keV $\gamma$-rays. The $p$-wave resonance was fitted including the background and $s$-wave component with the Breit-Wigner function and $1/v$ components after the normalization by the incident neutron-beam spectrum. The fitting result and obtained fitting parameters are shown in Fig.~\ref{Fitreso} and Table~\ref{resopara}, respectively. 
The broadening of the resonance shape derived from the thermal motion of gas atoms and the pulse shape of the neutron beam was taken into account in the fit. For the detailed formula of the fitting function, see APPENDIX (C),(D),and (E) in Ref.~\cite{okuda18}. Because the neutron width of the $p$-wave resonance is negligibly smaller than the $\gamma$ width of the $p$-wave resonance, the total width of the $p$-wave resonance was used as the $\gamma$-ray width of the $p$-wave resonance. \\

\subsection{Angular Distribution}\label{Ang}
\begin{figure}[htbp]
	\centering
	\includegraphics[width=0.9\linewidth]{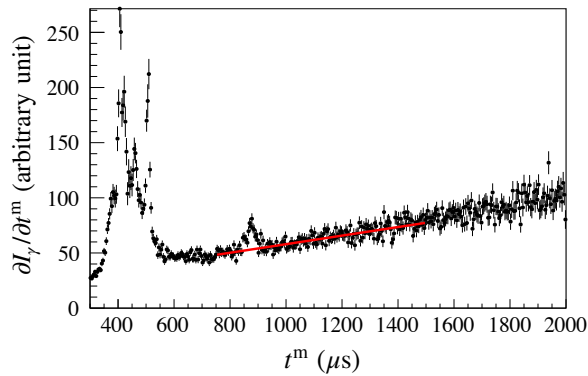}
	\caption[]{
	Histogram of $\partial I_{\gamma}/\partial t^{\rm m}$ gated with the full absorption peak of 7132~keV $\gamma$-rays for the single germanium crystal. The solid line shows the fit result of the $s$-wave and background components.   
	}
	\label{TOFGated_Fit}
\end{figure}

\begin{figure*}[htbp]
	\begin{center}
	\includegraphics[width=0.8\linewidth]{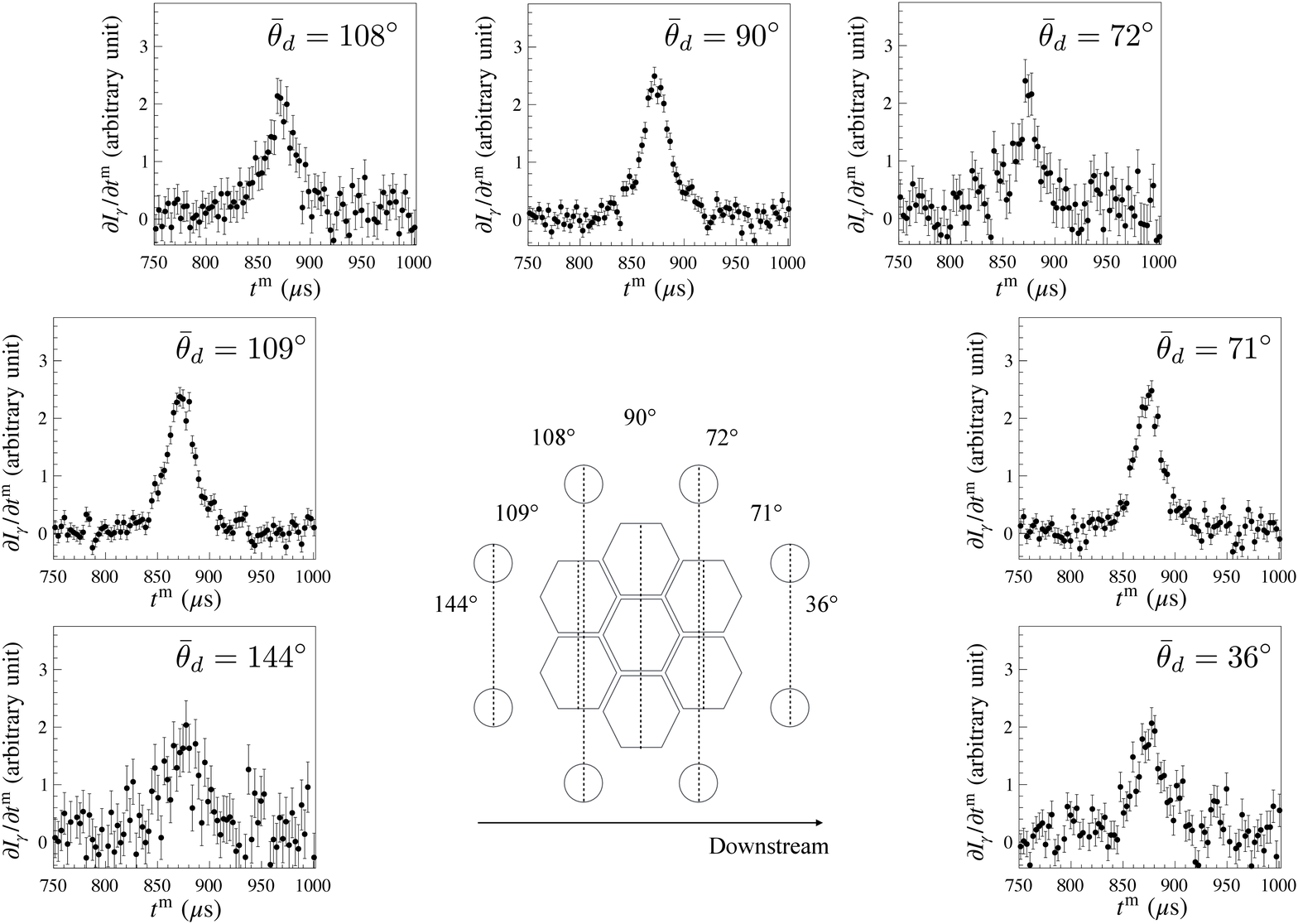}
	\caption[]{Histograms of $\partial I_{\gamma}/\partial t^{\rm m}$ gated with the single escape peak of 7132~keV $\gamma$-rays. The angles of the detector are denoted on the right top in each histograms.}
	\label{PwaveAngle}
	\end{center}
\end{figure*}

The interference of the $s$-wave and $p$-wave amplitudes results in an asymmetric shape of the $p$-wave resonance, which is dependent on the angle, as has been observed in case of $^{140}$La.~\cite{okuda18}. The background component needs to be subtracted before the evaluation of the shape of the $p$-wave resonance. Since the neutron absorption reaction of $^{56}$Fe does not have neutron resonances for $E_{\rm{n}}<$1keV, the 3.2~eV $p$-wave resonance of $^{131}$Xe+$n$ was isolated from the $s$-wave component of $^{131}$Xe+$n$ and a smooth background with a fit using $f(t^{\rm{m}})=at^{\rm{m}}+b$ with free parameters $a$ and $b$. For example, a histogram of  $\partial I_{\gamma}/\partial t^{\rm m}$ gated with the full absorption peak of 7132~keV after normalization of the incident neutron beam and fitting result are shown in Fig.~\ref{TOFGated_Fit}. The region of the $p$-wave resonance was excluded in the fitting. The $p$-wave resonance after the subtraction of the $s$-wave and background components for each detector angle is shown in Fig.~\ref{PwaveAngle}. \\

\begin{figure}[htbp]
	\begin{center}
	\includegraphics[width=1\linewidth]{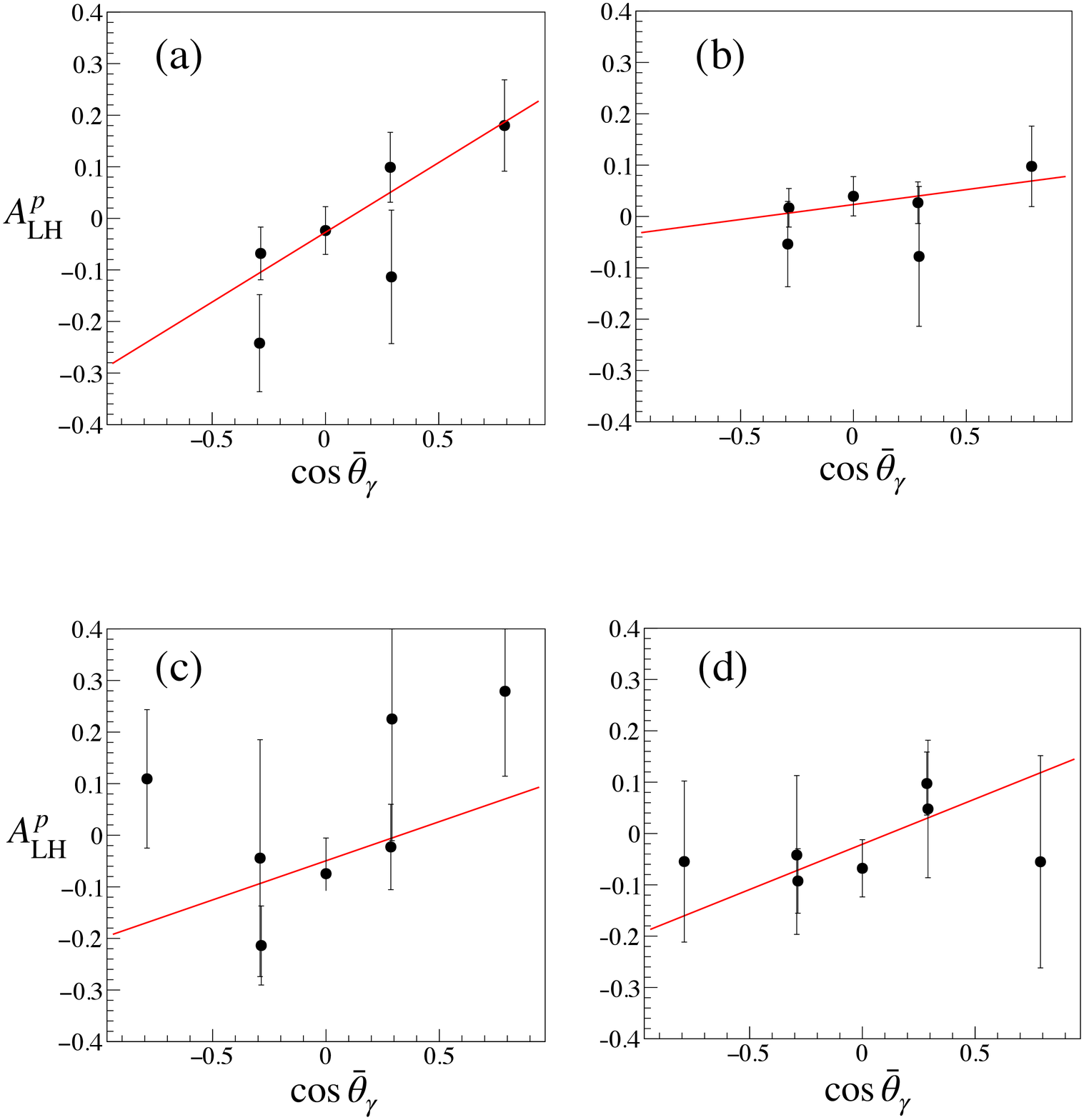}
	\caption[]{Angular distributions of $A^p_{\rm{LH}}$ measured with the full absorption peak and single escape peak using the 37.5~mm and 50.0~mm thick lead filters. (a), (b), (c), and (d) indicate the full absorption peak with the 37.5~mm thick lead filter, the single escape peak with the 37.5~mm thick lead filter, the full absorption peak with the 50.0~mm thick lead filter, and the single escape peak with the 50.0~mm thick lead filter, respectively. In the graphs of (a) and (b), the detectors at $144^\circ$ were not used in the fit because of the large dead time.}
	\label{A_02}
	\end{center}
\end{figure}

\begin{figure}[htbp]
	\centering
	\includegraphics[width=0.8\linewidth]{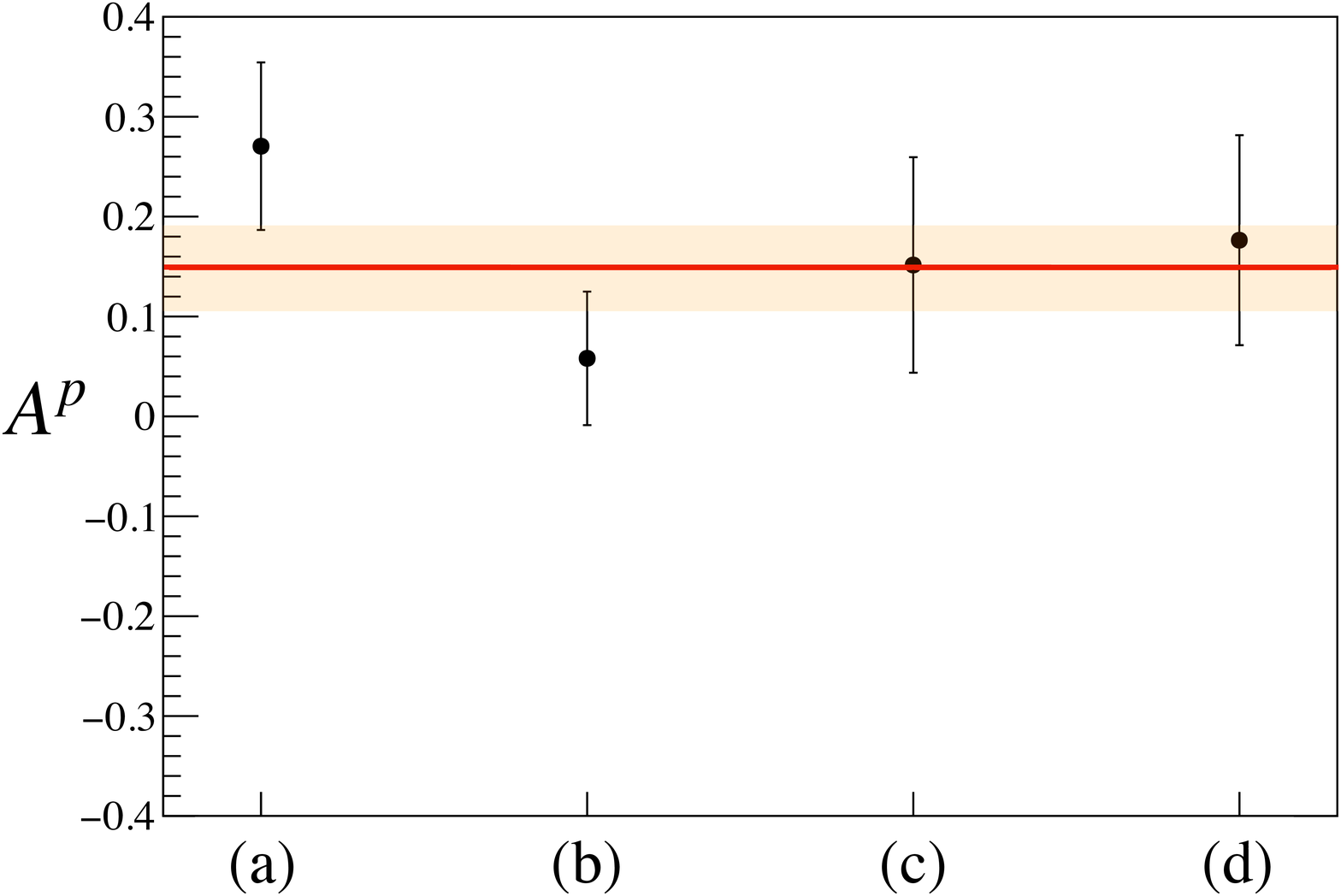}
	\caption[]{The values of the angular dependence $A^p$ for each condition. The solid line and colored region show the average value and 1$\sigma$ region, respectively. }
	\label{A_ave}
\end{figure}

The angular distribution of the $p$-wave resonance shape is measured by a "low-high asymmetry" defined as
\begin{eqnarray}
A^{p}_{\rm{LH}}(\theta_d)=\frac{N^p_{\rm L}(\theta_d)-N^p_{\rm H}(\theta_d)}{N^p_{\rm L}(\theta_d)+N^p_{\rm H}(\theta_d)},
\end{eqnarray}
where $\theta_d$ is the emission angle with respect to the incident neutron momentum, and $N^p_{\rm L}$ and $N^p_{\rm H}$ are integrals in the region of  $E_{\rm 2}-1.5\Gamma_{\rm 2} \le E_{n} \le E_{\rm 2}$ and $E_{\rm 2}\le E_{n} \le E_{\rm 2}+1.5\Gamma_{\rm 2}$, respectively. Variables $E_{\rm 2}$ and $\Gamma_{\rm 2}$ denote the resonance energy and total width of the $p$-wave resonance, which is defined by the $\gamma$ width and neutron width shown in Table~\ref{resopara} as $\Gamma_{\rm 2}=\Gamma^\gamma_2+\Gamma^n_2$. The detailed definitions of $N^p_{\rm L}$ and $N^p_{\rm H}$ are shown in Eq.~(8) in Ref.~\cite{okuda18} and Fig.~14 in Ref.~\cite{okuda18}. Note that $A^{p}_{\rm{LH}}(\theta_d)$, $N^p_{\rm L}$ and  $N^p_{\rm H}$ are calculated only for the $p$-wave resonance. 
\\The low-high asymmetry is plotted against the effective detector angle $\bar{\theta}_d$, which is obtained with a simulation of the germanium detector assembly~\cite{Takada18}, and fitted using a function of $A^p_{\rm {LH}}(\bar{\theta}_{d})=A^p\cos\bar{\theta}_d+B^p$ with free parameters $A^p$ and $B^p$. The angular distributions of $A^p_{\rm{LH}}$ are obtained under the following four conditions: (1) the full absorption peak with the 37.5 mm thick lead filter, (2) the single escape peak with the 37.5 mm thick lead filter, (3) the full absorption peak with the 50.0 mm thick lead filter, (4) the single escape peak with the 50.0 mm thick lead filter, as shown in Fig.~\ref{A_02}. The slope parameters $A^p$, which corresponds to the angular dependence of $p$-wave resonance shape, are plotted for the four conditions in Fig.~\ref{A_ave}. These results are consistent with each other, and their average value was $A^p=0.148\pm0.043$. The present result of the angular distribution is considered as an evidence of the interference between $s$- and $p$-wave amplitudes as already found in $^{140}$La and $^{118}$Sn~\cite{okuda18, yama20, okuda21, koga22}.

\section{Conclusion}
The $\gamma$-rays from the $^{131}$Xe($n$, $\gamma$)$^{132}$Xe reaction, especially for the 3.2~eV $p$-wave resonance, was studied with high statistics by using the enriched $^{131}$Xe target and filter. The $\gamma$-transition  from the $p$-wave resonance state to the $3^+$ state  at 1804~keV was clearly observed as well as already known $\gamma$-transitions in $^{132}$Xe.  The neutron-energy dependent angular distribution was evaluated with the low-high asymmetry using the direct transition, and slope parameter $A^p$ was determined to be 0.148$\pm$0.043.  This result will be interpreted by combining with the results of $^{140}$La and $^{118}$Sn in terms of the $s$-p mixing model for a more detailed understanding of the enhancement mechanism of the symmetry violation. The combined analysis will be reported in a separate paper.

\begin{acknowledgments}
The authors would like to thank the staff of beamline04 for the maintenance of the germanium detectors, and MLF and J-PARC for operating the accelerators and the neutron production target.  The neutron scattering experiment was approved by the Neutron Scattering Program Advisory Committee of IMSS and KEK (Proposals Nos. 2014S03, 2015S12, 2018S12). The neutron experiment at the Materials and Life Science Experimental Facility of the J-PARC was performed under a user program (Proposal Nos. 2019A0113 , 2020B0356, 2020B0451). This work was supported by JSPS KAKENHI Grant Nos. 17H02889, and the US National Science Foundation PHY-1913789. W. M. Snow acknowledges support from the Indiana University Center for Spacetime Symmetries.
\end{acknowledgments}
\bibliography{ngamma}
\end{document}